\documentclass[10pt,letterpaper]{article}
\usepackage{opex3}
\usepackage{geometry}
\usepackage{color}
\usepackage{graphicx}
\usepackage{mathptmx}
\usepackage{courier}
\usepackage{helvet}
\usepackage{sidecap}
\begin{document}




\title{Phase-coherent detection of an optical dipole force by Doppler velocimetry}

\author{M. J. Biercuk$^{1,2}$, H. Uys$^{1,3}$, J. W. Britton$^{1}$,\\ A. P. VanDevender$^{1}$, and
 J. J. Bollinger$^{1}$}
\address{$^{1}$National Institute of Standards and Technology, Ion Storage Group, Boulder, CO 80305\\
$^{2}$School of Physics, The University of Sydney, NSW 2006 Australia\\
$^{3}$National Laser Centre, Council of Scientific and Industrial Research, Pretoria, South Africa}

\email{michael.biercuk@sydney.edu.au} 

\homepage{http://www.nist.gov/pml/div688/grp10/index.cfm}
\homepage{http://physics.usyd.edu.au/$\sim$mbiercuk} 


\begin{abstract} We report phase-coherent Doppler detection of optical dipole forces using large ion crystals in a Penning trap.  The technique is based on laser Doppler velocimetry using a cycling transition in $^{9}$Be$^{+}$ near 313 nm and the center-of-mass (COM) ion motional mode. The optical dipole force is tuned to excite the COM mode, and measurements of photon arrival times synchronized with the excitation potential show oscillations with a period commensurate with the COM motional frequency.  Experimental results compare well with a quantitative model for a driven harmonic oscillator.  This technique permits characterization of motional modes in ion crystals; the measurement of both frequency and phase information relative to the driving force is a key enabling capability -- comparable to lockin detection -- providing access to a parameter that is typically not available in time-averaged measurements. This additional information facilitates discrimination of nearly degenerate motional modes.   \end{abstract}



\section{Introduction}
\indent In quantum informatic and quantum control settings, trapped ions have demonstrated their versatility for high-fidelity studies of entanglement~\cite{Turchette98, Sackett2000, Blatt8Ion, Reichle2006, Blatt2008, Monroe09}, coherent control~\cite{Bible, BiercukQIC2009, Monroe_Fast2010}, quantum logic~\cite{Monroe95, BlattCZ, Leibfried2003, Monroe_Transverse09}, precision spectroscopy~\cite{Schmidt2005}, and quantum simulation~\cite{QSim98, Leibfried2002, BlattDirac, Monroe_QSim}.  In addition to the manipulation of internal atomic states, these experiments rely on the coherent excitation of normal motional modes of these charged particles in approximately harmonic (pseudo-)potential wells~\cite{Bible}.  The Coulomb interaction between neighboring ions allows common motional modes to serve as a quantum coherent bus for sharing information between particles, and has functioned as a fundamental mechanism for the experimental realization of two-qubit gates~\cite{Leibfried2003} and multipartite entanglement~\cite{Turchette98}.  Recently, exploitation of transverse motional modes in linear ion crystals has been proposed as a means by which large-scale quantum information processing may be realized using trapped ions~\cite{CiracPorras_SpinBoson, MonroeTransverse}.
\\
\indent Implementing a multi-qubit quantum logic operation requires the accumulation of a conditional phase between constituent particles.  In trapped-ion experiments, extant techniques rely on the ability to apply a state-dependent force to a Coulomb crystal of harmonically bound trapped ions.  Such a force may be realized using an optical dipole excitation created by a pair of off-resonant excitation laser beams.  In short,  the application of a state-dependent ac Stark shift to trapped ions in a spatially graded light field results in the ions experiencing differential forces dependent upon their internal quantum states (e.g. for a qubit $F_{\uparrow}$ and $F_{\downarrow}$).  This provides a means to conditionally excite normal modes of motion.  In the interaction frame the excited harmonic oscillator coherently acquires a state-dependent phase.  Applying this interaction for an appropriate duration leads to a maximal entanglement of the internal states of multiple ions, and can be used to implement a multipartite quantum logic operation~\cite{Leibfried2003}.  
\\
\indent Exploiting ion motional modes in this way requires precise knowledge of motional frequencies, mode spacings, absolute and relative force strengths, and motional coherence.  Such characterization is generally accomplished via measurement of time-averaged ion fluorescence under continuous or pulsed excitation of motional modes, and the observation of motional sidebands.  This technique is highly efficient for small ion crystals~\cite{King98} in which only a few motional modes exist, and where the optical-dipole interaction is weakly sensitive to the alignment of the excitation-beam interference wavefronts relative to a particular crystallographic direction.  However, such detection becomes inefficient and complex in large crystals~\cite{BlattCrystals, mitchell98, Bollinger2000} with densely packed mode structures, and in which the Lamb-Dicke parameter makes observation of sidebands difficult.
\\
\indent We are thus motivated to develop methods by which ion motional excitation may be observed directly, and individual motional excitations may be resolved in a densely packed mode structure.  The experiments we present in this manuscript describe the detection of optical dipole excitation of trapped-ion motional modes via laser Doppler velocimetry~\cite{mitchellDoppler, Berkeland1998}.  We utilize phase-coherent detection of the transverse center-of-mass (COM) motional mode of a two-dimensional crystal of $\sim$100 $^{9}$Be$^{+}$ ions in a Penning trap~\cite{Biercuk_yN_2010}, excited using state-dependent optical dipole forces.  Experimental measurements of the temporal modulation of photon arrival times agree well with theoretical predictions for a driven harmonic oscillator and provide both amplitude and phase information about the excited motion.
\\
\indent The remainder of this paper is organized as follows.  In Sec.~\ref{Sec:Expt} we briefly introduce our experimental system, followed by a detailed description of phase-coherent motional excitation and detection via Doppler velocimetry in Sec.~\ref{Sec:Technique}.  Sec.~\ref{Sec:OD} describes our excitation-laser system for generation of optical dipole forces and presents phase-coherent Doppler detection measurements. We move on to a study of the benefits of maintaining phase information in the discrimination of nearly degenerate motional modes in Sec.~\ref{Sec:Phase}, and conclude with a discussion and a future outlook.

\section{\label{Sec:Expt}Experimental system}
\indent The technique we describe is general and may be applied in RF Paul or Penning traps;  here we focus on experimental implementation in a macroscopic Penning trap.  This is a device allowing for charged-particle confinement in three dimensions using static electric and magnetic fields in an ultra-high-vacuum envelope.  The trap consists of a stack of axially aligned cylindrical electrodes in the presence of a $\sim$4.5 T, axially oriented external magnetic field.  Axial confinement for charged particles is achieved by the application of a static electric potential difference between endcap and center electrodes (here $\sim$1,200 V, see Fig.~\ref{Fig:Raman}c).  Radial confinement is realized by ion motion in the axially oriented magnetic field producing a centripetal Lorentz force~\cite{Bollinger1984,brel88}.
\\
\indent Ions are Doppler laser-cooled to an axial temperature $\sim$0.5 mK \cite{jenm04, jenm05} using $\sim$313-nm UV laser light red-detuned from an atomic transition between the $2s$ $^{2}S_{1/2}$ and $2p$ $^{2}P_{3/2}$ manifolds of $^{9}$Be$^{+}$. When cooled, the ions form arrays with well defined crystal structure \cite{itaw98, mitchell98, jenm04, jenm05} and ion spacing $\sim$15 $\mu$m  due to Coulomb repulsion.  The ion array rotates rigidly at a frequency $\sim$50 kHz due to the geometry of the confining electric and magnetic fields, and the rotation frequency is precisely controlled by an external rotating dipole potential \cite{huap98a, huap98b}.  Adjusting the rotation rate of the array allows controllable manipulation of the crystal aspect ratio and dimensionality (planar vs. 3D).  For these experiments we focus on crystals with $\sim$100 ions in a two-dimensional planar array having a  diameter $\sim$300 $\mu$m.
\\
\indent Quartz windows allow optical access to the trap along the axial direction (parallel to $\vec{B}$) and also transverse to the applied magnetic field (the ``side-view'').  An $f/5$ imaging system connected to a CCD camera or phototube (selectable via an electrically controlled flipper mirror) allows for direct side-view imaging of resonantly scattered photons from the ion crystal or detection of the total resonant fluorescence.  Pulsed control over all laser beams is provided by switching the rf-drive applied to acousto-optic modulators (AOMs). The time-domain control system is based on a programmable logic device under computer control.

\section{\label{Sec:Technique}Phase-coherent detection of ion motional modes}
\begin{figure}[tb]
\centering\includegraphics[width=10cm]{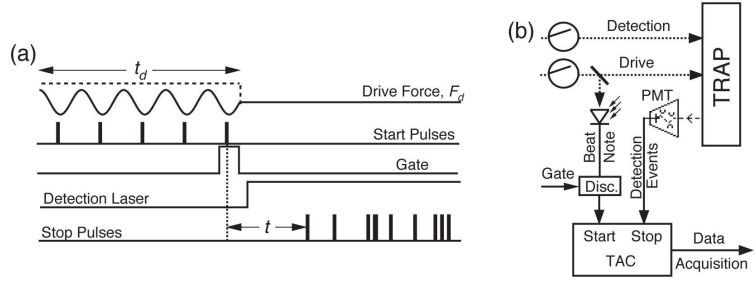}
\caption{\label{Fig:Schematic}Schematic of excitation and detection scheme.  (a) Details of pulse-timing used in phase-coherent excitation and detection of motional drive. (b) Schematic of experimental measurement apparatus.  Dashed lines represent optical signals, solid lines represent electrical signals.  Switches controlled via TTL signals shown schematically in panel (a).}
\end{figure}

\indent In an approximately harmonic confining potential, ions experience shared normal-modes of motion, and $n$ ions exhibit $3n$ normal modes.   In our experiments we will aim to excite a normal mode of motion using an optical dipole force, and detect this excitation directly.  A variety of three-dimensional motional modes in a one-component $^{9}$Be$^{+}$ plasma have been studied previously~\cite{mitchellDoppler} in this system.  In this work we focus exclusively on the center-of-mass (COM) motional mode transverse to the crystal plane, where ions undergo uniform axial motion parallel to the magnetic field.  For our operating parameters the COM-mode resonance frequency is $\sim$867 kHz, although over a timescale of hours it drifts downward as heavy-mass ions (e.g. BeH$^{+}$) accumulate in the trap due to background-gas collisions.
\\
\indent In previous work we demonstrated phase-coherent detection of ion motion in which the electrically excited motional response was synchronized to the external driving force.    It was shown that this technique was capable of discriminating motional responses to forces at the scale of yoctonewtons.  A detailed characterization of the method, its sensitivity, performance limits, and scaling with external parameters is presented in~\cite{Biercuk_yN_2010} and the associated supplementary material.  Here we present only a review of the salient elements of the technique.
\\
\indent Motional detection is accomplished using laser Doppler velocimetry.  In this technique, ion motion in a direction parallel to the propagation direction of a laser beam modulates the intensity of resonant fluorescence due to Doppler shifts.  In these experiments we employ the axially directed laser used for Doppler cooling near 313 nm as a probe of ion motion in the confining potential of the Penning trap.  The laser is red-detuned from the Doppler cooling cycling transition by $\sim$12 MHz such that its frequency sits near the steepest section of the atomic profile mentioned above, where Doppler cooling efficiency and detection sensitivity are maximized~\cite{mitchellDoppler} (Fig. 1a).  Under oscillatory ion motion, the intensity of ion fluorescence will exhibit periodicity with the same frequency as the ion motion.
\\
\indent  A pulsed force is applied for a fixed duration, after which the Doppler detection light is switched on and resonantly scattered photons are detected using a phototube (~\ref{Fig:Schematic}).  The driving force, $F_{d}$, is applied at a fixed frequency, and a voltage discriminator is used to produce a chain of ``start'' pulses (here, from a photodiode interference signal) synchronized to the drive that are then sent to a time-to-amplitude converter (TAC).  The TAC is gated such that it accepts only the final start pulse prior to the end of the driving force pulse.  A ``stop'' pulse is generated by the detection of a resonantly scattered photon on the phototube, and is fed into the TAC.  The voltage output of the TAC scales linearly with the difference between start and stop pulse arrival times, and is read using a triggered data acquisition board.  The experiment is repeated a fixed number of times, and statistics of photon arrival times synchronized to the drive are collected.
\\
\indent In order to understand the expected signal in a Doppler velocimetry experiment we begin by presenting a quantitative analysis of the driven harmonic oscillator. Considering uniform ion motion in a quadratic potential we may write the equation of motion for the axial COM mode in terms of the equation of motion for a single ion as
\begin{equation}
m(\ddot{z}+\omega_{z}^{2}z)=F_{d}\sin(\omega_{d} t + \psi),
\end{equation}
where $m$ is the ion mass, $z$ is the axial coordinate, $\omega_{z}$ is the resonant oscillation frequency, $F_{d}$ is the magnitude of the spatially uniform driving force, and $\omega_{d}$ and $\psi$ are the frequency and the phase of the drive, respectively.  Without loss of generality we may set $\psi\equiv0$, and assuming $(\omega_{z}-\omega_{d})/\omega_{z}\ll1$ and $z(0)=\dot{z}(0)=0$, we find 
\begin{equation}
z(t)\approx\frac{2F_{d}}{m(\omega^{2}_{z}-\omega^{2}_{d})}\cdot\sin\left[\frac{1}{2}\left(\omega_{d}-\omega_{z}\right)t\right]\cdot\cos\left[\omega_{z}t+\frac{1}{2}\left(\omega_{d}-\omega_{z}\right)t\right].
\end{equation}
After application of a driving force at frequency $\omega_{d}$ of duration $t_{d}$ an ion will undergo a steady-state sinusoidal oscillation with velocity
\begin{equation}
\label{Eq:velocity1}
\dot{z}(t)=v\sin\left[\omega_{z}t+\phi\right],
\end{equation}

\noindent where the amplitude $v$ is defined by
\begin{equation}
\label{Eq:velocity2}
v=\frac{2F_{d}\omega_{d}}{m(\omega_{z}^{2}-\omega_{d}^{2})}\sin\left[\frac{(\omega_{z}-\omega_{d})t_{d}}{2}\right],
\end{equation}

\noindent and the oscillation phase $\phi$ is
\begin{equation}
\label{Eq:velocity3}
\phi=\frac{(\omega_{d}-\omega_{z})t_{d}}{2}.
\end{equation}

\noindent Doppler detection directly probes the ion velocity, as described in equations~\ref{Eq:velocity1}-~\ref{Eq:velocity3}.  The measurement protocol we have developed preserves the phase information of the driven harmonic oscillator (Eq.~\ref{Eq:velocity3}) rather than simply averaging measurements over random oscillation phases.

\section{\label{Sec:OD}Direct observation of the optical dipole force}
\begin{figure}[tb]
\centering\includegraphics[width=10cm]{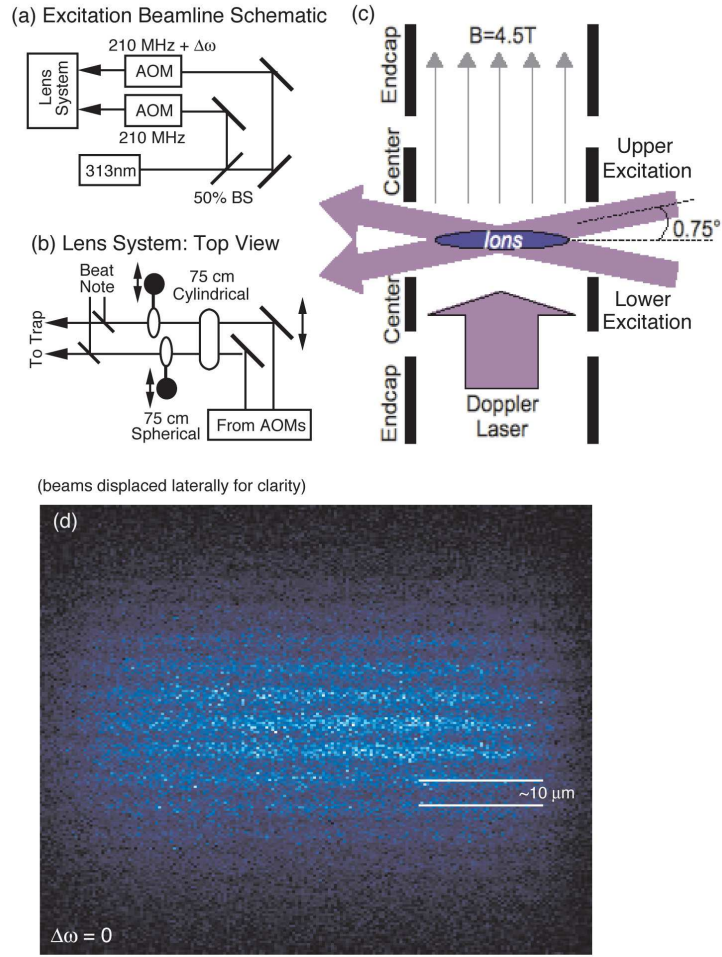}
\caption{\label{Fig:Raman}Laser excitation setup.  (a) Schematic diagram of laser beamline.  (b) Schematic of lens system used for creation of oblate beams with independent control of wavefront tilts.  Double-headed arrows indicate translation stages.  In experimental setup beams are stacked vertically, and converge with crossing angle of 1.5$^{\circ}$.  (c) Schematic side-view of laser excitation beamline showing trap electrodes, ion cloud, and orientation of axial magnetic field.  (d) Side-view image of fluorescence from the excitation lasers when tuned to be degenerate ($\Delta\omega=0$) and on-resonance with cycling transition used for Doppler detection.  The stationary interference pattern arising from lasers used to excite the optical dipole force is clearly visible.  Here, the ions are heated above the liquid-solid phase transition and do now show crystalline order.}
\end{figure}

\indent Optical dipole forces are applied using a two-color laser setup~\cite{Bible} that produces a traveling-wave interference pattern.  By appropriately tuning the laser frequency and polarization (relative to the level structure of the trapped ions), the laser can produce ac Stark shifts of varying magnitude and sign.  The presence of a spatial gradient in the light field then causes the ions to experience a force as they seek regions of low or high electric field intensity, depending on the sign of the Stark shift.
\\
\indent Our experimental setup involves splitting a 313 nm laser beam into two beamlines, each passing through an independently controlled AOM (Fig.~\ref{Fig:Raman}a).  One beam is frequency shifted by 210 MHz and the other by 210 MHz + $\Delta\omega$.  The beams enter the trap stacked vertically, and nearly perpendicular to the applied magnetic field at a shallow angle of $\theta=\pm0.75^{\circ}$, intersecting at the location of the planar ion crystal (Fig.~\ref{Fig:Raman}c).  Before entering the trap each beam passes through a $f=$ 75 cm cylindrical lens ($f$ indicates focal length) and a $f=$ 75-cm spherical lens to produce an oblate beam with a $\sim$10:1 aspect ratio, and a minimum waist parallel to the trap axis of $\sim$80 $\mu$m at the location of the ion crystal (Fig.~\ref{Fig:Raman}b).  Each beam is coarsely aligned using a large cloud of bright ions illuminated by the cooling beam while the excitation-laser beams are tuned to optically pump the ions to a dark state.  Adjusting the positions of the independently controlled 75-cm spherical lenses allows for maximization of the beam overlap with the ion crystal.  Care is taken to ensure that the beams enter the trap in the same vertical plane.  

\indent When $\Delta\omega=0$ a standing-wave interference pattern is established where the beams intersect.  We verify that the beams are striking the ions and that they are overlapped by direct imaging; the beams are tuned near the Doppler-cooling cycling transition and we image the standing-wave interference pattern as alternating bright and dark bands on an ion cloud above the liquid-solid phase transition~\cite{Bollinger1984, Bollinger1994}.  For the small angle at which the beams cross and the $\sim$313 nm wavelength we observe interference fringes with spacing $\sim$10 $\mu$m (Fig.~\ref{Fig:Raman}d).  The phase wavefronts of the interference pattern are aligned with the trap axes through beam steering and the addition of a mirror mounted on a linear translation stage which can laterally displace the lower beam relative to the upper beam before each passes through the 75-cm lenses (Fig.~\ref{Fig:Raman}b).  A combination of translation of the mirror with translation of the spherical lens allows for adjustment of the wavefront angles relative to the ion crystal. 
\\
 \indent Optical-dipole excitation of the COM mode is realized when $\Delta\omega\equiv\omega_{d}\approx\omega_{Z}$.  In this situation, the overlapped beams produce a ``running wave'' interference pattern, where the AC electric field amplitude at a fixed location oscillates temporally at $\Delta\omega$~\cite{Bible}.  Equation 1 is valid if the axial extent of a crystallized plane of ions is small relative to the wavelength of the interference pattern (the Lamb-Dicke criterion).  Here, the Lamb-Dicke parameter for a single ion at $T\approx 1$ mK is 
 \begin{equation}
 \eta=\sqrt{\frac{\hbar}{2m_{Be}\omega_{Z}}}2k\sin(\theta)\sqrt{\overline{n}+1}\approx0.07
 \end{equation}
where $k$ is the wavevector of the composite two-color laser system, and the average thermal phonon occupation, $\overline{n}=23$.

\begin{figure}
\centering\includegraphics[width=8cm]{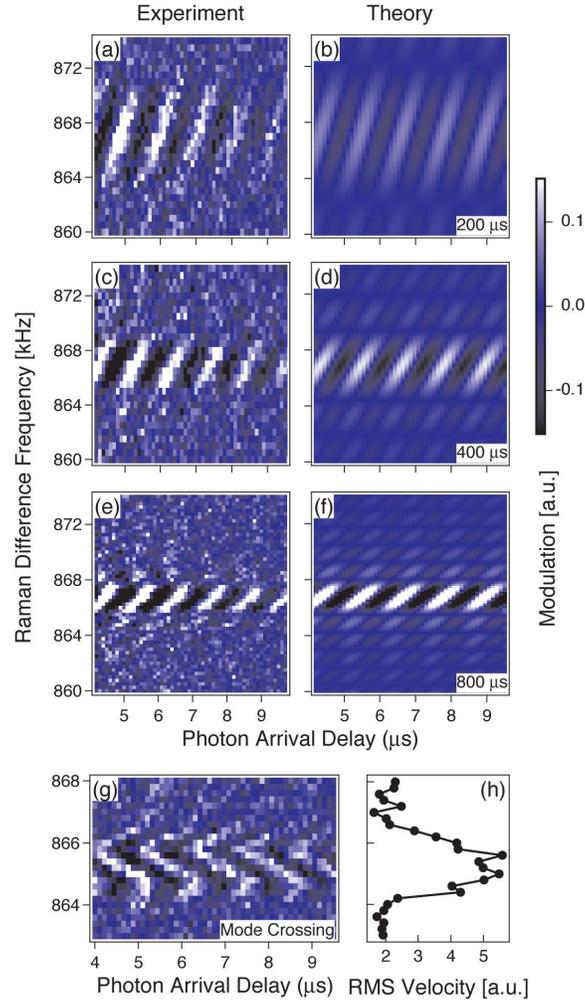}
\caption{\label{Fig:Doppler}Direct detection of the optical dipole force in a two-dimensional ion crystal.  (a,c,e) Experimental measurements of phase-coherent Doppler velocimetry as a function of excitation-laser difference frequency (vertical) and time delay for arrival of first detected photon.  Times shorter than $\sim$4.5 $\mu$s are discarded due to hardware delays.  Color scale indicates deviation from a fit to a simple exponential accounting for photon arrival statistics.  (b,d,f) Theoretical calculations using equations 4 and 5, and a fit to the magnitude of the oscillation amplitude.  Theory replicates salient experimental features including the resonance-linewidth scaling with drive time, and phase evolution with drive frequency. (g) Optical dipole force excitation of a mode nearly degenerate with the COM.  In this measurement the phase evolution across resonance changes abruptly indicating the presence of multiple motional resonances.  (h) Integrated root-mean-squared velocity of the resonance, using coherent excitation and detection, but averaging over phase information.}
\end{figure}

\indent We set the laser frequency $\sim$3 GHz red of the cooling/detection cycling transition, laser polarization $\sigma^{+}$, and $\Delta\omega\approx\omega_{Z}$.  A beat-note is generated for the detection system by picking off $\sim$5$\%$ of the excitation beam power immediately before the beams enter the trap, and interfering the beams on a photodiode.  This beat-note is used to synchronize photon arrival times with the drive excitation.  In this configuration the beat-note is stable on a timescale of up to a few seconds before undergoing $\sim$$\pi$ phase-slips due to interferometric instability between the two excitation laser beam paths.
\\
\indent We measure the arrival time of the first scattered photon, synchronized to the excitation beat-note, as a function of $\Delta\omega$.  In the data presented in Fig.~\ref{Fig:Doppler} the colorscale represents the residual, in the presence of the dipole excitation, of an exponential fit to photon arrival times used to account for photon arrival statistics in the absence of optical-dipole excitation~\cite{Biercuk_yN_2010}.  Modulation of the Doppler velocimetry signal due to ion motion is manifested as positive or negative deviations in the detected photon number at a particular time from the exponential background.  Periodic modulations are thus represented by alternating dark and light bands as a function of delay time from the beat trigger.
\\
\indent Experimental data match well with theoretical calculations based on the equations presented above (Fig.~\ref{Fig:Doppler}), showing a clear resonance feature near $\omega_{Z}/2\pi=867$ kHz.  Doppler velocimetry indicates a linear phase shift in the oscillator as $\omega_{d}$ is tuned through resonance, and  the linewidth of the resonance scales as the inverse of $t_{d}$.  The excitation of the oscillator is zero when $|\omega_{d}-\omega_{z}|=2\pi/t_{d}\times n$, for $n$ an integer, producing oscillation sidebands.  At excitation nulls, the external force and system response desynchronize and resynchronize~\cite{Leibfried2003}, yielding a closed loop in position-momentum phase-space.  We see sidelobes of the main resonance; however, these are in general difficult to observe, possibly due to timing jitter induced by interferometric instability in producing a beat-note.
\\
\indent In the experimental data we also observe a damping of the oscillation strength as a function of the delay time due to radiation pressure from the detection laser.  This effect is not accounted for in the calculations shown in the second column of Fig.~\ref{Fig:Doppler}.  In these calculations the strength of the applied force is used as a model parameter, and is extracted from theoretical fits to the oscillation amplitude as a function of drive frequency, as in previous experiments.  Because the strength of the drive force is extracted from the average oscillation amplitude over the whole oscillation period, the failure to account for laser-induced damping provides a slight underestimate of the optical dipole force strength.  However, we clearly observe motional coherence beyond 10 $\mu$s.
\\
\indent Calibration of the Doppler velocimetry signal relative to a given applied force is accomplished using quasistatic electric fields as described in detail in~\cite{Biercuk_yN_2010}.  The maximum observed force as deduced from theoretical fits to the measured data and comparison against the calibration described above is approximately 100 yN for $\sim$3 GHz detuning from resonance and power of $\sim$5 mW/beam.  This measured force strength matches well to theoretical calculations of the achievable force given the excitation beam detuning, configuration, polarization, and power~\cite{Bible}, within a factor of order unity (300 yN for these laser conditions).  The strength of the optical dipole force scales inversely with the detuning of the excitation laser frequency from the relevant atomic transition~\cite{Bible}, and we have verified this scaling qualitatively by choosing operating detunings from $\sim1-12$ GHz.  

 \indent Using this detection technique we are able to maximize the strength of the measured COM excitation by precisely aligning the excitation laser beams relative to the planar ion crystal through an iterative search in the beam positions.  The magnitude of the observed Doppler signal is sensitive not only to misalignment of the center of each beam's waist relative to the ion cloud (which reduces the average AC-Stark shift experienced by the ions), but also to misalignment of the interference wavefronts relative to the orientation of the crystal plane. In this setup, a lateral shift of the lower beam's mirror translator (coupled with a shift of the associated 75 cm spherical lens) relative to the upper beam by $\sim$200 $\mu$m tilts the interference wavefronts by 0.5$^{\circ}$ relative to the ion plane, resulting in a $\sim\pi/2$ phase shift of the force across the diameter of a crystal with $\sim$100 ions.  Spatially inhomogeneous force application results in reduced contrast in Doppler detection measurements due to spatial averaging arising from the ion rotation in the trap.  Thus, we are able to optimize the measured dipole force strength through micron-scale translations in the relative beam positions, aligning the interference wavefronts in two dimensions with precision much better than 0.5$^{\circ}$.
\\
\section{\label{Sec:Phase}Benefits of maintaining phase information}
\indent Typical measurements of motional modes use incoherent probes of, e.g. microwave energy absorption, to reveal resonant absorption features.  These approaches are limited by the fact that it is often difficult to discriminate between a single absorption resonance at frequency $\omega_{0}$ with Fourier-linewidth $t_{d}^{-1}$, and a pair of closely spaced resonances at frequencies  $\omega_{0}\pm \Delta/2$, with $\Delta$ the mode separation.  Resolving closely spaced resonance features using standard techniques requires increasing the interrogation time to produce a more favorable ratio of $\Delta/t_{d}^{-1}$, which may be undesirable for technical or practical reasons.
\\
\indent By contrast, coherent excitation and detection of closely spaced motional modes provides a means to improve discrimination by extracting information from the phase degree of freedom.  We have experimentally performed this kind of discrimination, identifying features in the phase degree of freedom that are obscured in averaged experiments.  In Fig.~\ref{Fig:Doppler}g we present measurements of Doppler velocimetry as a function of drive frequency that reveal a resonant feature with internal structure.  Instead of a simple monotonic evolution of the oscillation phase as the drive frequency is swept across resonance, we see a more complex pattern in which the phase evolution abruptly changes direction twice, producing a zig-zag pattern.  
\\
\indent We associate the observed complex phase evolution in the motional resonance with the crossing of a higher-order asymmetric mode with the symmetric COM mode.  Plasma-mode calculations show that for a planar ion-crystal asymmetric, drum-head, or tilt modes can arise in which motion in the axial direction is not uniform across all ions.  Asymmetric modes are Doppler shifted by the rotation of the planar ion crystal and, for certain rotation frequencies, produce apparent mode crossings in the lab frame.  
\\
\indent This particular mode-crossing evolved in time from an independent spectral feature to an apparent crossing, possibly due to the gradual accumulation of heavy-mass ions that centrifugally separate from $^{9}$Be$^{+}$.  A time-varying fraction of heavy-mass ions produces a Doppler-shifted mode whose resonant frequency shifts in the lab frame.  Further, this feature appeared in a circumstance where, prior to the accumulation of a significant fraction of heavy-mass ions, the Doppler velocimetry signal was weak, indicating poor alignment of the excitation-laser interference wavefronts with the ion crystal.
\\
\indent In order to further investigate the benefits of this kind of phase-sensitive detection in discriminating nearly degenerate motional modes, we calculate the velocimetry response for an excitation of duration $t_{d}$ at frequency $\omega_{d}$, in the presence of two modes at $\omega_{0}\pm \Delta/2$, with $\Delta$ the mode separation (Fig. 4).  The first row of this figure compares the integrated RMS velocimetry signal using phase-coherent excitation and detection against an incoherent excitation in which total energy absorption is measured.  For large $\Delta$, the individual resonances are easily resolved by either technique.  However, as $\Delta\rightarrow^{+}t_{d}^{-1}$, energy absorption measurements produce a ``double-hump'' feature that eventually becomes a single peak with a non-Lorenzian lineshape for $\Delta<t_{d}^{-1}$.  In some regimes both techniques produce a double-hump, but phase-coherent excitation and detection produces a deeper trough due to interference between the closely spaced modes (Fig. 4d).
\\
\indent Additional information is provided by directly examining the phase degree of freedom in a coherent excitation and detection, either as a function of time (second row, Fig.4) or only at $t=0$ (third row).  Maintaining this information provides an additional means to distinguish between closely spaced resonance features.  For sufficiently small $\Delta$, distinguishability is provided by interference between the sidebands (Fig. 4g) producing a pattern distinct from the single-peak feature (Fig. 4f).  Alternatively, by adding a temporal offset between the conclusion of the drive period and the beginning of the velocimetry measurement (bottom row, Fig. 4), one may allow the closely spaced resonances to acquire phase $\sim\pi$, producing destructive interference.  The ability to employ these approaches to extracting additional information from phase-coherent doppler velocimetry are predominantly limited by the $Q$ of the mechanical resonance, and clear observation of sidebands.  Additionally, as the required time delay increases as $\Delta^{-1}$, the time required to resolve closely spaced resonances must be traded-off against the benefits of using an increased drive time, reducing the Fourier linewidth of the resonances.

\begin{figure}[htbp]
\centering\includegraphics[width=12cm]{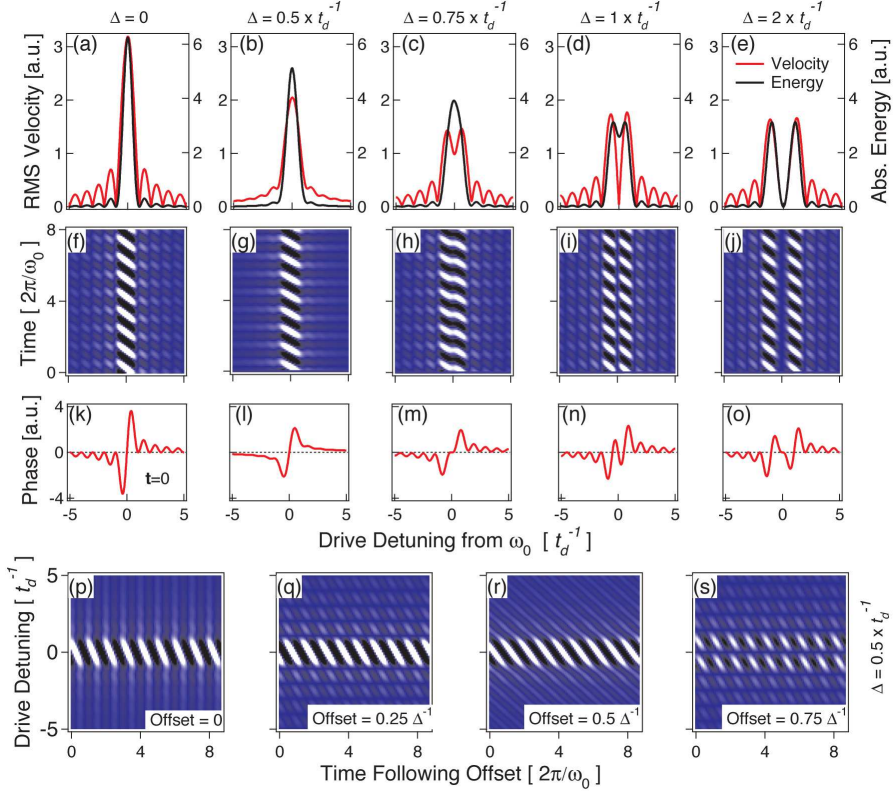}
\caption{\label{Fig:Phase}Comparison of calculated phase-coherent Doppler velocimetry to energy-absorption measurments in the presence of closely spaced motional modes.  (a-e) Integrated RMS velocity and absorbed energy for motional modes with spacings, $\Delta$, designated by the column headers, and parameterized by $t_{d}^{-1}$. (f-j) Two-dimensional doppler velocimetry signal as a function of drive detuning (horizontal) and photon arrival time (vertical).  Phase information in both the central resonances and sidebands due to interference between closely spaced motional modes permits distinguishability relative to energy absorption approaches. (k-o) One-dimensional phase-trace demonstrating information contained in the phase-degree of freedom for $t=0$.  (p-s) Demonstration of interference between closely spaced motional modes due to the inclusion of a delay ("Offset") between conclusion of the drive and commencement of Doppler detection, presented in units of the inverse peak spacing $\Delta^{-1}$.  For a fixed value of $\Delta$, accumulation of a time-dependent phase shift between motional modes improves resolving power.}  
\end{figure}

\section{Discussion and conclusion}
\indent Optical dipole forces are widely employed in atomic physics, and have become a fundamental tool in the realization of quantum gates for trapped-ion quantum information.  In these experiments, optical dipole forces are tuned to provide a force which is conditioned on the internal state of a trapped atomic ion, and as such produce a means by which information may be coherently transferred from an internal to an external degree of freedom~\cite{Bible}.  Precise characterization of motional modes is therefore required in order to realize a robust quantum operation.  In the case of ion pairs or small ion crystals only a small number of motional modes are present and the system is not particularly sensitive to the alignment of the excitation lasers relative to the crystallographic axes.  In such settings a simple measurement of ion fluorescence as a function of excitation frequency is sufficient to characterize the optical dipole excitation of a given motional mode.  However, as ion trap quantum information experiments move towards more complex and larger ion crystals, mode densities increase and it becomes increasingly difficult to distinguish particular resonant frequencies using these techniques.  Further, alignment of the excitation lasers with the crystal becomes more important as the size of a one- or two-dimensional crystal increases.  With these considerations in mind, it becomes clear that a diagnostic technique providing more information about the optical dipole force becomes important as we look to scale-up ion trap experiments.
\\
\indent In this manuscript we have detailed the phase-coherent excitation and detection of motional modes in a $\sim$100-ion crystal via optical dipole forces.  This technique has provided a direct means to observe the optical dipole force using laser Doppler velocimetry, and provided quantitative information about the magnitude of the driving force, the phase and frequency of the excited motional mode, and the effect of radiation damping.   
\\
\indent   Our results demonstrate that it is possible to exploit the phase information in our detection technique to discriminate nearly degenerate motional modes spaced by less than the Fourier-limited linewidth of a particular resonance.  This capability is especially important in large crystals where the mode density is high and it may be difficult to distinguish between narrowly spaced modes using time-averaged measurements in which phase-information is discarded.  In addition we have shown that it is possible to observe asymmetric motional excitations and mode-crossings using this technique, and that it may be employed as a sensitive probe of beam alignment relative to an ion crystal. 
\\
\indent In these experiments we have focused on excitation of the COM mode of a two-dimensional ion crystal using a single detection beam oriented perpendicular to the ion plane.  In this configuration we are limited to the detection of motion along the detection beam.  However, it is in principle possible to expand this system by performing sequential Doppler velocimetry measurements with a series of detection beams oriented along orthogonal directions.  Similarly, the addition of a spatially resolving top-view imaging system allows time-resolved detection of ion fluorescence from individual ions, and hence the ability to directly detect general nonsymmetric drumhead modes.  We believe that this technique will serve as a useful diagnostic tool for laser-beam alignment and motional-mode characterization in complex experiments using large multidimensional ion crystals.
\\
\\
\noindent This manuscript is a contribution of NIST, a US government agency, and is not subject to copyright.  This work was supported by the DARPA OLE program and by IARPA.  M.J.B.  acknowledges partial fellowship support from IARPA and GTRI.  The authors wish to thank C. Ospelkaus for technical assistance, James Chou and Brian Sawyer for comments on the manuscript, and T. Rosenband for valuable discussions.

\end{document}